\documentclass[a4paper, amsfonts, amssymb, amsmath, reprint, showkeys, nofootinbib, twoside, superscriptaddress]{revtex4-1}
\usepackage[english]{babel}
\usepackage[utf8]{inputenc}
\usepackage{comment}
\usepackage{array}
\usepackage[acronym]{glossaries}
\usepackage{physics}
\usepackage{bm}
\usepackage{mathtools}
\usepackage{placeins}
\usepackage{enumitem}

\begin{document}
\title{Simulating Vibrational Dynamics on Bosonic Quantum Devices}

\author{Shreyas Malpathak}
\affiliation{Department of Physical and Environmental Sciences,
University of Toronto Scarborough, Toronto, Ontario M1C 1A4, Canada}
\affiliation{Chemical Physics Theory Group, Department of Chemistry,
University of Toronto, Toronto, Ontario M5S 3H6, Canada}
\author{Sangeeth Das Kallullathil}
\affiliation{Department of Physical and Environmental Sciences,
University of Toronto Scarborough, Toronto, Ontario M1C 1A4, Canada}
\affiliation{Chemical Physics Theory Group, Department of Chemistry,
University of Toronto, Toronto, Ontario M5S 3H6, Canada}
\author{Artur F. Izmaylov}
\affiliation{Department of Physical and Environmental Sciences,
University of Toronto Scarborough, Toronto, Ontario M1C 1A4, Canada}
\affiliation{Chemical Physics Theory Group, Department of Chemistry,
University of Toronto, Toronto, Ontario M5S 3H6, Canada}
\date{\today}

\begin{abstract}
Bosonic quantum devices, which utilize harmonic oscillator modes to encode information, are emerging as a promising alternative to conventional qubit-based quantum devices, especially for the simulation of vibrational dynamics and spectroscopy. We present a framework for digital quantum simulation of vibrational dynamics under anharmonic potentials on these bosonic devices. In our approach the vibrational Hamiltonian is decomposed into solvable fragments that can be used for Hamiltonian simulation on currently available bosonic hardware. Specifically, we extended the Cartan subalgebra approach [T.C. Yen, A.F. Izmaylov, PRX Quantum 2, 040320 (2021)] -- a method for decomposing quantum Hamiltonians into solvable parts -- to bosonic operators, enabling us to 
construct anharmonic Hamiltonian fragments that can be efficiently diagonalized using Bogoliubov transforms. The approach is tested using a simulation of tunneling dynamics in a model two-dimensional double-well potential and calculations of vibrational eigenenergies for small molecules. Our fragmentation scheme provides a new approach for digital quantum simulations on bosonic quantum hardware for multi-mode anharmonic vibrational dynamics. 

\end{abstract}

\maketitle
\section{Introduction}
Quantum algorithms for quantum chemistry on qubit-based quantum devices have seen tremendous development over the last decade \cite{kassal2011,bauer2020,mcardle2020,ollitrault2021}. Although algorithms for electronic structure theory have been at the forefront of these advances, calculation of vibrational and vibronic spectra have also recently started gaining attention \cite{mcardle2019,sawaya2019,ollitrault2020,sawaya2020,sawaya2021,lee2021,magann2021,majland2023,richerme2023,trenev2023,loaiza2024b}. However, the latter algorithms require a boson-to-qubit mapping that necessitates truncating the bosonic Fock space above a chosen threshold \cite{mcardle2019,sawaya2020}. Moreover, the bosonic unitary transforms that appear in these problems need to be mapped onto qubit-based unitary gates, which presents an extra overhead cost \cite{liu2024}. 

Bosonic quantum devices have emerged as a promising alternative to conventional qubit-based devices \cite{liu2024,crane2024,dutta2024a,dutta2024b,kang2024,araz2024}. They feature bosonic degrees of freedom (also referred to as `qumodes') coupled to qubits, and are thus often referred to as hybrid continuous variable (CV) - discrete variable (DV) devices. Exploiting the coupling between CV and DV degrees of freedom, these devices can perform arbitrary bosonic unitary transforms on qumodes using both Gaussian and non-Gaussian bosonic unitary gates \cite{liu2024}. The availability of bosonic gates makes these devices well suited for vibrational and vibronic problems, obviating the need for boson-to-qubit mapping. Numerous studies for the computation of vibrational \cite{sparrow2018} and vibronic \cite{huh2015,huh2017,hu2018,clements2018,shen2018,wang2020,jahangiri2020,jnane2021,wang2022,macdonell2023} spectra, and chemical dynamics \cite{macdonell2021,lyu2023,wang2023,cabral2024,de_albornoz2024,so2024,sun2024} have already been undertaken. They vary in the type of bosonic architecture used and the digital versus analog nature of the simulation.
Recently, in an analog simulation of \textit{anharmonic} chemical dynamics, the Hamiltonian for a one-dimensional quartic double well was mapped onto the Hamiltonian of a bosonic device to calculate tunneling rate constants \cite{cabral2024}. However, this analog approach is limited by the device Hamiltonians available, and does not exploit the universality offered by digital simulations. Capturing anharmonic dynamics using digital simulations is difficult as it requires the use of non-Gaussian gates. Owing to this difficulty, to the best of our knowledge, a general approach for digital quantum simulation for anharmonic vibrational dynamics is missing and sorely needed. To bridge this gap, we introduce a novel fragmentation-based approach that can be implemented on currently available bosonic devices.


Our approach relies on writing the vibrational Hamiltonian as a sum over \textit{solvable} fragments,
\begin{align}
    H = \sum_k H_k = \sum_k \mathcal{U}_{k}\,D_k \, {\mathcal{U}_k}^{\dagger}.
\end{align}
The solvable fragments $H_k = \mathcal{U}_{k}\,D_k \, {\mathcal{U}_k}^{\dagger}$, where $D_k$ are diagonal, are specially constructed so that they can be easily diagonalized using gates of bosonic quantum devices. In qubit-based quantum computing for electronic structure theory, such Hamiltonian fragmentation algorithms have found success  for optimizing the number of measurements \cite{yen2023,choi2023a,choi2023b,patel2024a,patel2024b}, reducing errors in Hamiltonian simulations using Trotter product formulas \cite{martinez2023,martinez2024}, and for linear combination of unitaries (LCU) based approaches \cite{loaiza2023a,loaiza2023b,loaiza2024a,patel2024c}. To construct solvable fragments for the vibrational Hamiltonian we extend the Cartan sub-algebra approach \cite{yen2021} to bosonic operators. This allows us to construct quartic Hamiltonian fragments that can be diagonalized using Bogoliubov transforms, with possible generalizations to higher orders. These solvable fragments can be used with the Trotter product formula to approximate the propagator as, 
\begin{align}
    e^{-iHt} \approx \prod_{k} e^{-iH_kt } = \prod_{k} {\mathcal{U}_k}\, e^{-i D_k t }\, {\mathcal{U}_k}^{\dagger}.
\end{align}  
The error from the Trotter approximation can be estimated using existing methods \cite{childs2021,martinez2023,rendon2024,martinez2024}, and made smaller than the required error threshold by choosing a sufficiently small Trotter time step. Taken together, our fragmentation procedure allows for the implementation of the propagator of the vibrational Hamiltonian for applications to vibrational dynamics and spectra.

The rest of the paper is organized as follows. Section~\ref{sec:theory} lays out the theory to construct solvable fragments for the vibrational Hamiltonian, including details about Bogoliubov transforms, the form of the diagonal fragments, their implementation using bosonic gates, and an  algorithm to find solvable fragments using a non-linear optimization procedure. In Sec.~\ref{sec:results} we demonstrate our fragmentation approach for two applications:  tunneling dynamics in a two dimensional double-well potential and calculating vibrational eigenenergies of small molecules. Section~\ref{sec:conc} concludes.

\section{Theory} \label{sec:theory}
The vibrational Hamiltonian written in dimensionless normal coordinates $\bm{q}$ and their corresponding conjugate momenta $\bm{p}$ is \cite{mcardle2019},
\begin{align}
    {H} & = \frac{1}{2}\sum_{i=1}^N \omega_i\left( p_i^2 + q_i^2 \right)  + \sum_{i,j,k=1}^NV^{\{3\}}_{ijk} q_i q_j q_k \notag \\ 
    & + \sum_{i,j,k,l=1}^NV^{\{4\}}_{ijkl} q_i q_j q_k q_l. \label{eq:H_vib}
\end{align}
Here, we use atomic units with $\hbar = 1$, and have truncated the Taylor expansion of the vibrational potential energy surface $V(\bm{q})$ to fourth order around the equilibrium geometry. $N$ is number of vibrational degrees of freedom, and $V^{\{3\}}_{ijk}$ and $V^{\{4\}}_{ijkl}$ are cubic and quartic coefficients respectively. The position and momentum operators follow canonical commutation relations $[q_j,q_k] = [p_j,p_k] = 0$, and $[q_j,p_k] = i\delta_{jk}$. The vibrational Hamiltonian can also be expressed as a function of bosonic creation and annihilation operators,
\begin{align}
    b_j &= \frac{1}{\sqrt{2}}(q_j + i p_j),  & b_j^{\dagger} &= \frac{1}{\sqrt{2}}(q_j - i p_j)
\end{align}
that follow canonical commutation relations $[b_i,{b}_j^{\dagger}] = \delta_{ij}$, $[b_i,{b}_j] = 0$, and $[b_i^{\dagger},{b}_j^{\dagger}]$ = 0. In this form, due to the truncation at fourth order in the Taylor expansion, the vibrational Hamiltonian contains up to quartic terms in creation and annihilation operators,
\begin{align}
    {H} & = \sum_{i=1}^N\omega_i\left( {b}_i^{\dagger}{b}_i + \frac{1}{2}\right) \notag \\
    & + \sum_{i,j,k=1}^N \frac{V^{\{3\}}_{ijk}}{2\sqrt{2}}\left({b}_i^{\dagger}+{b}_i\right)\left({b}_j^{\dagger}+{b}_j\right)\left({b}_k^{\dagger}+{b}_k\right) \notag  \\
    & + \sum_{i,j,k,l=1}^N \frac{V^{\{4\}}_{ijkl}}{4}\left({b}_i^{\dagger}+{b}_i\right)\left({b}_j^{\dagger}+{b}_j\right)\left({b}_k^{\dagger}+{b}_k\right)\left({b}_l^{\dagger}+{b}_l\right).
\end{align} 

Our strategy is to decompose this vibrational Hamiltonian into \textit{solvable} fragments $H = \sum_k H_k$ by extending the Cartan subalgebra (CSA) approach \cite{yen2021} to bosonic operators. As mentioned in the introduction, these fragments are called solvable because they can be brought into diagonal form, $H_k = \mathcal{U}_{k}\,D_k \, {\mathcal{U}_k}^{\dagger}$, using `simple' unitaries. We elaborate on two aspects here. First, the diagonal operators $D_k$ are polynomials of the Cartan subalgebra (CSA)  operators \cite{yen2021}, in this case the number operators $n_i = b_i^{\dagger}b_i$. These operators are diagonal and commute with each other, $[n_i,n_j] = 0$, allowing us to easily exponentiate $D_k$ to obtain the propagator. Second, the unitaries $\mathcal{U}_{k}$ are chosen to be Bogoliubov unitaries that are easy to implement on bosonic devices using Gaussian gates \cite{braunstein2005b,weedbrook2012} (details presented in Sec.~\ref{sec:frag_prop}). Since the vibrational Hamiltonians considered here are quartic, the fragments themselves need to have at least up to quartic terms. However, to understand our approach of constructing solvable quartic fragments it is instructive to first look at the simpler case of quadratic Hamiltonians. 

\subsection{Quadratic Bosonic Hamiltonians} \label{sec:quad_ham}

We consider bosonic Hamiltonians with up to quadratic terms,
\begin{align}
     H_{lq} & = \sum_{p,q = 1}^{N} A_{pq} {b}_p^{\dagger}{b}_q + \frac{1}{2}B_{pq}{b}_p^{\dagger}{b}_q^{\dagger} + \frac{1}{2}B^{*}_{pq}{b}_p{b}_q \notag \\
     & + \sum_{p=1}^{N} C_p b_p^{\dagger} + C_p^* b_p,
\end{align}
with $A = A^{\dagger}$ and $B = B^{T}$ to ensure hermiticity.
They can be diagonalized using Bogoliubov transforms as \cite{Bogoliubov2009},
\begin{align}
    H_{lq} = \mathcal{U}_{b} \left(\sum_{p=1}^{N} \epsilon_p \tilde{b}_p^{\dagger}\tilde{b}_p + K \bm{1} \right) \mathcal{U}_{b}^{\dagger} \label{eq:diag_Hlq}.
\end{align}
Here, $K$ is a constant and $\bm{1}$ is the identity operator. Details about the Lie algebraic properties of $H_{lq}$ can be found in Appendix~\ref{app:quad_diag}. A Bogoliubov transform mixes and displaces the original bosonic variables into a new set of bosonic variables that also satisfy the canonical commutation relations. Details of Bogoliubov transforms and the unitary operators $\mathcal{U}_{b}$ that implement them can be found in Appendix~\ref{app:bog_transf}. In the proceeding section, we construct fragments that have a similar structure as in Eq.~\eqref{eq:diag_Hlq}, but with a quartic diagonal form, to obtain solvable quartic fragments.


\subsection{Solvable Quartic Fragments for the Vibrational Hamiltonian} \label{sec:sol_frags}

Unlike quadratic Hamiltonians that were discussed in the previous section, general bosonic Hamiltonians with cubic, quartic and/or higher order terms can not be diagonalized using simple unitary transforms like Bogoliubov transforms. However, taking inspiration from the quadratic case, we can exploit Bogoliubov transforms to \textit{construct} solvable quartic Hamiltonian fragments of the form,
\begin{align}
    H_{k} & = \mathcal{U}_b^{(k)} \left(\sum_{p,q=1}^{N} \eta^{(k)}_{p,q} \tilde{n}_p \tilde{n}_q \right) {\mathcal{U}_{b}^{(k)}}^{\dagger} \label{eq:sol-bos-frags},
\end{align}
where $\eta_{pq}^{(k)}$ are parameters associated with the fragment, $\tilde{n}_p = \tilde{b}_p^{\dagger}\tilde{b}_p$ is the number operator for the Bogoliubov transformed bosonic operators $\tilde{\bm{\xi}} = {\mathcal{U}_{b}^{(k)}}^{\dagger}\,\bm{\xi}\, \mathcal{U}_{b}^{(k)} \equiv \bm{M}^{(k)}\bm{\xi} + \bm{\Gamma}^{(k)}$. The matrix $\bm{M}^{(k)}$ and the vector $\bm{\Gamma}^{(k)}$ parameterize the Bogoliubov transform for the $k^{th}$ fragment with parameters $\{\alpha_{pq}^{(k)},\beta_{pq}^{(k)},\gamma_p^{(k)}\}$. Refer to Appendix~\ref{app:bog_transf} for relevant details of the Bogoliubov transform. In the original basis, the solvable fragments in Eq.~\ref{eq:sol-bos-frags} will contain up to quartic terms in the original bosonic operators. To decompose the vibrational Hamiltonian into such solvable fragments, $H  = \sum_{k} H_{k}$,  a non-linear optimization over the parameters of the fragments $\{\eta_{pq}^{(k)},\alpha_{pq}^{(k)},\beta_{pq}^{(k)},\gamma_p^{(k)}\}$ can be performed to find the fragments. Here we use a greedy algorithm to find the fragments one at a time. For finding solvable two-electron fragments for the electronic structure Hamiltonian, this algorithm is referred to as the Greedy Full Rank Optimization (GFRO) \cite{yen2021}. The steps of the algorithm are as follows.
\begin{enumerate}

    \item Collect the coefficients of the cubic and quartic terms in the Hamiltonian, hereon referred to as $c_j$.

    \item \label{en:opt} Find a solvable fragment $H_k$ by optimizing the cost function $\sum_j (c_j -c_j^{(k)})^2$ over the fragment parameters $\{\eta_{pq}^{(k)},\alpha_{pq}^{(k)},\beta_{pq}^{(k)},\gamma_p^{(k)}\}$, where $c_j^{(k)}$ are the coefficients of the cubic and quartic terms of the solvable fragment written in the original basis. The fragment parameters are restricted to be real.

    \item \label{en:store} Store the solvable fragment $H_k$, and subtract it from the original Hamiltonian, $H \to H - H_k$. This has the effect of changing the cubic and quartic coefficients in the Hamiltonian as $c_j\to c_j - c_j^{k}$, while also modifying the linear, quadratic and constant terms.

    \item \label{en: tol} Repeat steps \ref{en:opt} and \ref{en:store} till the residual of the coefficients of the cubic and quartic terms $\sqrt{\sum_{j}  (c_j -\sum_{k=1}^{N_f} c_j^{(k)})^2}$ is less than a specified tolerance. For a sufficiently small tolerance, the remaining cubic and quartic terms can be discarded. $N_f$ is the number of quartic solvable fragments found.

    \item The remaining Hamiltonian, $H - \sum_{k=1}^{N_f} H_k$, only contains up to quadratic terms, and can be diagonalized using a Bogoliubov transform. This fragment is referred to as $H_0$. 
\end{enumerate}
At the end of the algorithm, the Hamiltonian is decomposed into $N_f +1$ solvable fragments $H = \sum_{k=0}^{N_f} H_k$, where $H_0$ is a quadratic fragment, and $H_k$ for $k \in [1,N_f]$ are quartic fragments.

\subsection{Constructing the Propagator from Hamiltonian Fragments}  \label{sec:frag_prop}

The decomposition of the vibrational Hamiltonian into solvable fragments can be used in conjunction with the Trotter approximation for the propagator, 
\begin{align}
    e^{-iHt} \approx \prod_{k=0}^{N_f} e^{-iH_kt} = \prod_{k=0}^{N_f} \mathcal{U}_b^{(k)} e^{-iD_kt}  {\mathcal{U}_{b}^{(k)}}^{\dagger},
\end{align}
where $D_0 = \sum_{p=1}^{N} \epsilon_p \tilde{n}_p + K \bm{1} $ is the diagonal quadratic fragment, and $D_k = \sum_{p,q=1}^{N} \eta^{(k)}_{p,q} \tilde{n}_p \tilde{n_q} $ are the diagonal quartic fragments for $k \in [1,N_f]$. To implement the propagator on a bosonic quantum device, the unitaries corresponding to the Bogoliubov transforms, $\mathcal{U}_{b}^{(k)}$, and the diagonal fragment unitaries, $e^{iD_kt}$, need to be decomposed into bosonic gates. Using the Bloch-Messiah decomposition of the symplectic Bogoliubov matrix $\bm{M}$ \cite{braunstein2005,houde2024}, the Bogoliubov unitary can be decomposed into elementary Gaussian unitary gates as follows, 
\begin{align}
    \mathcal{U}_b & = \left(\prod_{p=1}^{N} \mathcal{D}_p(\gamma_p)\right) \left(\prod_{p>q=1}^{N} BS_{p,q}\left(2\phi_{pq},\frac{\pi}{2}\right)\right) \notag \\
    & \times \left(\prod_{p=1}^{N} S_p\left(\zeta_p,0\right)\right) \left(\prod_{p>q=1}^{N} BS_{p,q}\left(2\chi_{pq},\frac{\pi}{2}\right)\right). \label{eq:bog_u_decomp}
\end{align}
The elementary Gaussian gates, displacement ($\mathcal{D}$), beam-splitter ($BS$), and one-mode squeezing ($S$) are defined as \cite{liu2024},
\begin{align}
    \mathcal{D}_p\left(\gamma\right) & = e^{\gamma b_p^{\dagger}- \gamma^{*} b_p}, \label{eq:u_disp}\\
    BS_{pq}\left(\theta,\phi\right)   & =  e^{-i\frac{\theta}{2}\left(e^{i\phi}b_p^{\dagger}b_q + e^{-i\phi}b_q^{\dagger}b_p\right)}, \label{eq:u_bs} \\
    S_p\left(\zeta\right) & =e^{ \frac{1}{2} \left(\zeta{b_p^{\dagger}}^2 -\zeta^{*} b_p^2\right)} \label{eq:u_sq}.
\end{align}
The subscripts for the gates denote the bosonic modes they operate on. Displacement and one-mode squeezing are single-mode gates, whereas beam-splitter is a two-mode gate. It is important to note that the displacement gate for the $p^{th}$ mode is denoted $\mathcal{D}_p$, which is distinct from the $p^{th}$ diagonal fragment $D_p$. Also note that although the parameters corresponding to displacement and one-mode squeezing gates can be complex, for the Bogoliubov transforms considered in this work we only require real parameters. Details about the Bloch-Messiah decomposition for the Bogoliubov matrix $\bm{M}$ and the relation between parameters $\gamma_p, \chi_{pq}, \zeta_{p},$ and $\phi_{pq}$ in Eq.~\eqref{eq:bog_u_decomp} to the parameters of the Bogoliubov transform are provided in Appendix~\ref{app:unr_bog}.

\begin{figure}
    \centering
    \includegraphics[width=0.4\textwidth]{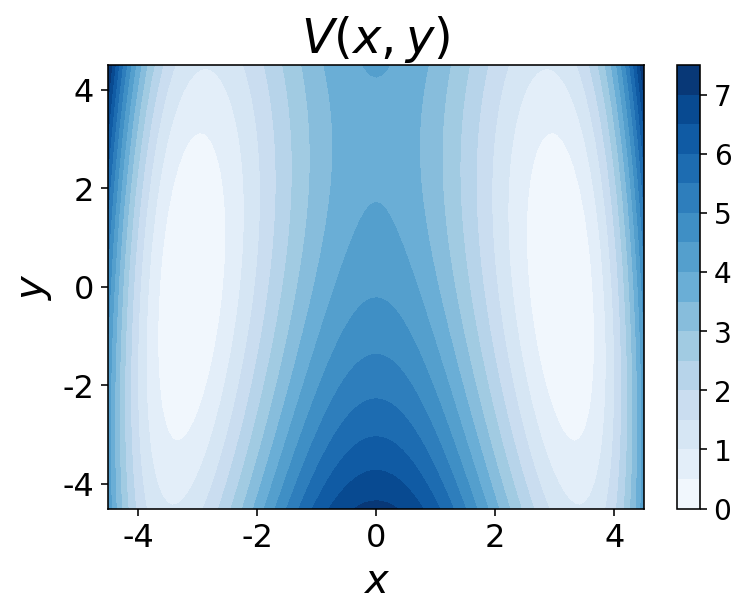}
    \caption{A contour plot of the two-dimensional double well potential $V(x,y)$ in units of $\hbar\omega_x$.}
    \label{fig:2dw}
\end{figure}

The unitary operators corresponding to the diagonal fragments can also be decomposed into gates available on bosonic hardware. The unitary $e^{iD_0 t}$ corresponding to the quadratic diagonal fragment $D_0$ can be written as a product of phase space rotation gates and a phase factor, 
\begin{align}
    e^{-iD_0 t}  & = e^{-i Kt} \prod_{p=1}^{N} e^{-i \epsilon_p \tilde{n}_p t}  = e^{-i Kt} \prod_{p=1}^{N} R_p(\epsilon_p t).
\end{align}
The phase space rotation gate over the $p^{th}$ mode is defined as $R_p(\theta) = e^{-i \theta  n_p}$. Lastly, the unitary operators $e^{iD_k t}$ corresponding to the quartic diagonal fragments $D_k$ can be partitioned into a product of Kerr interactions,
\begin{align}
    e^{-iD_k t}  & =  e^{-i \sum_{p,q=1}^{N} \eta^{(k)}_{p,q} \tilde{n}_p \tilde{n}_q t} = \prod_{p,q=1}^{N}  e^{-i \eta^{(k)}_{p,q} \tilde{n}_p \tilde{n}_q t}.
\end{align}
Here $e^{-i \eta^{(k)}_{p,p} \tilde{n}_p \tilde{n}_p t}$ is the Kerr gate for the $p^{th}$ mode and $e^{-i \eta^{(k)}_{p,q} \tilde{n}_p \tilde{n}_q t}$ is a cross-Kerr interaction between the $p^{th}$ and $q^{th}$ mode. Both of them can be implemented on bosonic devices \cite{liu2024, araz2024}.

\section{Results} \label{sec:results}

We apply our fragmentation approach to two applications: tunneling dynamics for a two-dimensional model system and for the calculation of vibrational eigenenergies of small molecules.

\begin{figure}
    \centering
    \includegraphics[width=0.45\textwidth]{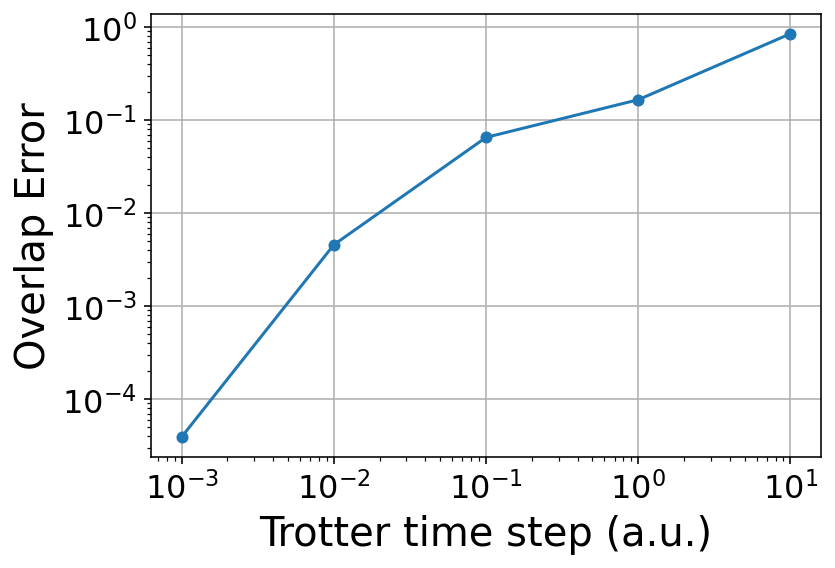}
    \caption{Overlap error, as defined in Eq.~\eqref{eq:ovlp_err}, as a function of the Trotter time step.}
    \label{fig:trott_err}
\end{figure}
\subsection{Tunneling Dynamics in Tropolone}

We study the proton transfer reaction in the excited state of Tropolone using a two-dimensional model based on \textit{ab-initio} calculations \cite{wojcik2009}. The model potential written in atomic units is,
\begin{align}
    V(x,y) & = \frac{\omega_x}{8x_0}\left(x - x_0\right)^2 \left(x +  x_0\right)^2 \notag \\
    & + \frac{\omega_y}{2}\left[y + \alpha \left(x^2 - x_0^2\right)\right]^2.
\end{align}
Here $x$ and $y$ are dimensionless coordinates representing the high-frequency proton tunneling mode and the nearly-planar hydrogen bond stretching mode respectively \cite{wojcik2009}. Figure~\ref{fig:2dw} depicts a contour plot of the potential. The parameters in the potential are $\omega_x = 3594$ cm$^{-1}$, $\omega_y = 414$ cm$^{-1}$, $x_0 = 3.15$ and $\alpha = 0.301$. Solvable fragments for this double well Hamiltonian are found using the GFRO algorithm presented in Sec.~\ref{sec:sol_frags}, with Powell's algorithm \cite{Powell1964} used for the non-linear optimization. A tolerance of $10^{-2}$ cm$^{-1}$ is used in step~\ref{en: tol} of the GFRO algorithm to stop fragmentation, resulting in a total of four solvable fragments including one quadratic fragment. 
We track tunneling dynamics of a Gaussian wavepacket initially centered at the bottom of the left well. It is created by displacing the vacuum state $\ket{\mathbf{0}}$ to be centered at the potential minimum located at $x = -x_0, y = 0$,
\begin{align}
    \ket{\psi(0)} \equiv \mathcal{D}_x(-x_0/\sqrt{2})\ket{\mathbf{0}}. \label{eq:psi0}
\end{align}
The average energy of this wavepacket is 4429 cm$^{-1}$, well below the barrier height of 14042 cm$^{-1}$. Owing to its  significant overlap with the ground and first excited eigenstates of the potential, upon evolution, the wavepacket coherently tunnels through the barrier on a timescale close to the period corresponding to the tunneling splitting between these two states ($\tau$). To construct the propagator, we find Hamiltonian fragments and use the Trotter approximation. Figure~\ref{fig:trott_err} presents the overlap error as a function of the Trotter time step, with 
\begin{align}
    \text{overlap error} = 1 - \left|\braket{\psi_{ex}(\tau)}{\psi_{tr}(\tau)}\right|, \label{eq:ovlp_err}
\end{align}
where $\ket{\psi_{ex}(\tau)}$ and $\ket{\psi_{tr}(\tau)}$ are the wavefunctions propagated exactly and using the Trotter approximation respectively. The trotter time step can be chosen to keep the overlap error below a chosen threshold. For larger systems, well established methods to estimate Trotter approximation errors can be used to obtain the appropriate step size \cite{childs2021,martinez2023,rendon2024,martinez2024}. Noting that the propagator is constructed from solvable fragments that can be implemented on a bosonic device, this demonstration offers a methodology to simulate tunneling dynamics on currently available devices.


\begin{table}
    \centering  
    \vspace{0.5cm}
  \begin{tabular}{|c|c|c|c|}
  \hline
    Molecule & Number of & Number of & Number of  \\
     & modes ($N$) & bosonic fragments & Pauli fragments  \\
    \hline 
    CO     & 1 & 2 & 54\\
    H$_2$O & 3 & 7 & 127 \\
    H$_2$S & 3 & 7 & 170 \\ 
    CO$_2$ & 4 & 4 & 54 \\ \hline
  \end{tabular}  
  \caption{The number of solvable fragments found for the various molecules considered in this work using two different fragmentation schemes: the bosonic fragmentation scheme presented in this work (bosonic fragments), fully commuting Pauli fragments of the vibrational Hamiltonian (Pauli fragments).}
  \label{tbl:num_frags}
\end{table}

\begin{table*}
    \centering
    \begin{tabular}{|c|c|c|c|c|c|c|c|c|c|}
    \hline
    Molecule & $n_{max}$ & \multicolumn{2}{c|}{Ground} & \multicolumn{2}{c|}{$1^{st}$ excited}  & \multicolumn{2}{c|}{$2^{nd}$ excited}  & \multicolumn{2}{c|}{$3^{rd}$ excited} \\
    \hline
    &  & Exact  & $H_{\rm{eff}}$ &  Exact  & $H_{\rm{eff}}$ & Exact  & $H_{\rm{eff}}$ & Exact  & $H_{\rm{eff}}$  \\ \hline
    CO      & 8 & 1213.59 & 1213.60 & 3624.80 & 3624.82 & 6017.79 & 6017.83 & 8395.91 & 8395.96 \\
    H$_2$O   & 6 & 4667.74  & 4667.82 & 6284.86 & 6284.93 & 7882.02 & 7882.09 & 8377.50 & 8377.61 \\    
    H$_2$S  & 7 & 3263.81 & 3263.83 & 4534.01 & 4534.04 &  5600.12 &  5600.13 & 5710.57 & 5710.58 \\
    CO$_2$  & 5 & 2532.06 & 2532.07 & 3186.21  &  3186.22 & 3186.21 & 3186.22 & 3822.36 & 3822.37 \\
    \hline
    \end{tabular}
    \caption{The first four vibrational eigenenergies for various molecules calculated using numerically exact diagonalization compared to those calculated using the Hamiltonian fragments along with the Trotter approximation for the propagator ($H_{\rm{eff}}$). All energies are presented in cm$^{-1}$. Also listed is the cutoff for the maximum occupation number $n_{max}$ used for each molecule.}
    \label{tbl:vib_eigs}
\end{table*}

\subsection{Vibrational Eigenenergies}

With the goal of finding the lowest vibrational eigenenergies of small molecules, we use our Hamiltonian fragmentation method to find solvable fragments for the following molecules: CO, H$_2$O, H$_2$S, CO$_2$. The vibrational Hamiltonians are constructed using harmonic frequencies and cubic and quartic coupling coefficients obtained from electronic structure calculations. For CO and H$_2$O we use the Gaussian package \cite{g16}, with HF/cc-pVDZ and B3LYP/cc-pVTZ  for CO and H$_2$O respectively. For CO$_2$ and H$_2$S we use the PySCF software \cite{pyscf} with HF/6-31G. The choice of the electronic structure method is not every important for the purpose of illustrating our fragmentation procedure. For future calculations where comparisons with benchmark results or experiments is required, more sophisticated electronic structure methods may be used. Moreover, only two-mode couplings have been considered here as they are easily available in the software packages used. However, our method can handle arbitrary mode couplings. The fragments are found using the GFRO algorithm presented in Sec.~\ref{sec:sol_frags} with the BFGS algorithm \cite{NoceWrig06} used to perform the non-linear optimization. A tolerance of $0.1$ cm$^{-1}$ is used in step~\ref{en: tol} of the GFRO algorithm to stop the fragmentation procedure.

Table~\ref{tbl:num_frags} lists the number of fragments found for the various molecules using two different fragmentation procedures: the bosonic fragmentation scheme proposed in this work and the fully commuting (FC) grouping of Pauli fragments that is commonly used for applications on qubit-based quantum algorithms \cite{yen2020}. We 
used the direct boson-to-qubit mapping \cite{somma2003} to map the vibrational Hamiltonian to qubits, with the cutoff for the maximum occupation number $n_{max}$ listed in Table.~\ref{tbl:vib_eigs}. We use the sorted insertion algorithm \cite{crawford2021} to find FC fragments. The number of bosonic fragments is found to be $13-27$ times smaller than the number of FC fragments. Although the cost of the full simulation also depends on the unitaries that rotate the fragments into their respective diagonal bases and the Trotter approximation error for the chosen fragmentation scheme, having fewer fragments certainly lowers the cost of the simulation. Thus, the bosonic fragmentation scheme for bosonic quantum devices proposed here promises to be a competitive alternative to conventional qubit-based approaches for calculating vibrational eigenenergies.

To calculate vibrational eigenenergies, the propagator is constructed using the Trotter approximation, 
\begin{align}
    e^{-iHt} \approx \prod_{k=0}^{N_f} e^{-iH_kt} \equiv e^{-iH_{\rm{eff}}t}.
\end{align}
To demonstrate our approach, we calculate the eigenenergies of $H_{\rm{eff}}$ for the small molecules considered here. The maximal evolution time is generally decided by the accuracy required in the eigenvalue calculation. Furthermore, the total evolution will need to be partitioned into smaller time steps to ensure that the Trotter approximation error is maintained below the acceptable error threshold. Here, for the purpose of demonstrating our fragmentation scheme, we use an evolution time of $t = 1$ a.u., which is sufficiently small to obtain Trotter approximation errors in eigenvalues $<1$ cm$^{-1}$. Other methods based on perturbative estimates for the Trotter approximation error for eigenvalues may also be employed \cite{martinez2024}. Table~\ref{tbl:vib_eigs} compares the first four eigenenergies for the molecules considered here obtained using the Trotter approximation to those obtained by the exact diagonalization of the Hamiltonian. For all molecules considered, the fragmentation procedure yields errors $< 1$ cm$^{-1}$ compared to the exact results.

\section{Conclusion} \label{sec:conc}

Going beyond the state-of-the-art for simulations of the vibrational problem on bosonic devices, we have introduced a novel approach for digital simulation of vibrational dynamics for general anharmonic vibrational Hamiltonians. Our approach relies on partitioning  the Hamiltonian into solvable fragments, allowing us to construct the time evolution operator on bosonic hardware using the propagators for individual fragments with the Trotter approximation.  Solvable anharmonic fragments are constructed by rotating diagonal fragments using Bogoliubov transforms. Although restricted to quartic Hamiltonians in this study, the approach can be further extended to higher orders by constructing the solvable fragments as higher-order polynomials of bosonic number operators. 

The fragmentation scheme is tested on two applications. We perform coherent tunneling dynamics for a two-dimensional double-well potential. Noting that this model Hamiltonian can be partitioned into just four solvable fragments, this system may serve as a promising candidate for an experimental demonstration of our approach in future studies. Furthermore, the ability of the approach to fragment Hamiltonians with higher number of  degrees of freedom is demonstrated by calculating vibrational eigenenergies for molecules with up to four vibrational modes. A comparison of the number of bosonic fragments with the number of fully commuting Pauli fragments highlights that the bosonic fragmentation scheme is $13 - 17$ times cheaper in terms of fragment counts. Furthermore, we wish to emphasize that for systems where a more accurate description of the potential energy surface is required, our fragmentation scheme can also be modified to the vibrational Hamiltonian written in the second quantized Christiansen form \cite{christiansen2004} that is based on the $n-$mode representation of the potential \cite{bowman2003}. Our approach extends the range of applicability of bosonic quantum devices to simulations of chemical dynamics and spectroscopy.

\section*{Acknowledgments}
The authors thank Dr. Stepan Fomichev and Dr. Ignacio Loaiza for providing the Hamiltonians of H$_2$S and CO$_2$ molecules. S.M. also thanks Prof. Nathan Wiebe, Smik Patel and Dr. Alexey Uvarov for helpful discussions. This work was funded under the NSERC Quantum Alliance program. This research was partly enabled by Compute Ontario (computeontario.ca) and the Digital Research Alliance of Canada (alliancecan.ca) support. Part of the computations were performed on the Niagara supercomputer at the SciNet HPC Consortium. SciNet is funded by Innovation, Science, and Economic Development Canada, the Digital Research Alliance of Canada, the Ontario Research Fund: Research Excellence, and the University of Toronto.

\bibliographystyle{unsrt}

\bibliography{bibfile}

\appendix 

\section{Lie algebraic properties of quadratic bosonic Hamiltonians} \label{app:quad_diag}

A general bosonic Hamiltonian over $N$ modes with only quadratic terms can be written as,
\begin{align}
    H_{quad} = \sum_{p,q = 1}^{N} A_{pq} {b}_p^{\dagger}{b}_q + \frac{1}{2}B_{pq}{b}_p^{\dagger}{b}_q^{\dagger} + \frac{1}{2}B^{*}_{pq}{b}_p{b}_q,
\end{align}
with $A = A^{\dagger}$ and $B = B^{T}$ to ensure hermiticity. We note that the operators $\{{b}_p^{\dagger}{b}_q + \frac{1}{2}\delta_{p,q}, {b}_p^{\dagger}{b}_q^{\dagger},{b}_p{b}_q \}_{p,q=1}^{N}$ form $2 N^2 + N$ distinct generators of the Lie algebra $\mathfrak{sp}(2N,\mathrm{C})$ \cite{arvind1995,guaita2024}. However, linear combinations of these generators with complex coefficients do not necessarily form Hermitian operators. Instead, $H_{quad}$ can be expressed as 
\begin{align}
    H_{quad} = \sum_{k} c_k X_k,
\end{align}
where $c_k$ are real coefficients and $X_k$ are the hermitian generators of the compact symplectic Lie algebra $\mathfrak{sp}(2N)$ and can be constructed as hermitian linear combinations of the generators of $\mathfrak{sp}(2N,C)$. The maximal tori theorem for compact groups \cite{yen2021,hall} then guarantees that $H_{quad}$ can be diagonalized using a unitary from the corresponding Lie group, the compact symplectic group SP$(2N)$. Furthermore, this can be extended to Hamiltonians with quadratic and linear terms, 
\begin{align}
    H_{lq} =  H_{quad} + \sum_p C_p b_p^{\dagger} + C_p^* b_p.
\end{align}
For such Hamiltonians, we note that the operators $\{\mathrm{1},b_p,b_p^{\dagger}\}_{p=1}^N$ along with the quadratic bosonic operators span the affine symplectic Lie algebra $I\mathfrak{sp}(2N,\mathrm{C})$ \cite{ferraro2005}. Similar arguments as above can then be used to see that $H_{lq}$ can be diagonalized with corresponding Lie group unitaries. As discussed in Sec.~\ref{sec:quad_ham} these are Bogoliubov unitaries. In the Cartan subalgebra (CSA) method \cite{yen2021}, one approach is to use Lie group unitaries to find solvable fragments. This is the approach we employ here, and is detailed in Sec.~\ref{sec:sol_frags}.

\section{Details of the Bogoliubov Transform} \label{app:bog_transf}

As mentioned in the main text, Bogoliubov transforms mix and displace the original bosonic modes $\bm{\xi}$ into new bosonic modes $ \tilde{\bm{\xi}}$, defined as,
\begin{align}
    \left(\begin{array}{c}
    \tilde{\bm{b}} \\
    \tilde{\bm{b}}^{\dagger}  \\
    \end{array}\right) = \underbrace{\left(\begin{array}{cc}
    \bm{U} & -\bm{V} \\
    -\bm{V}^{*}  &  \bm{U}^{*} \\
    \end{array}\right)}_{\bm{M}} \left(\begin{array}{c}
    \bm{b} \\
    \bm{b}^{\dagger}  \\
    \end{array}\right) 
    +  \underbrace{\left(\begin{array}{c}
    \bm{\gamma} \\
    \bm{\gamma^{*}}  \\
    \end{array}\right)}_{\bm{\Gamma}}. \label{eq:bog_transf}
\end{align}
Here  $\bm{\xi} \equiv \left(\bm{b},\bm{b}^{\dagger}\right)^T,$ and $\tilde{\bm{\xi}} \equiv \left(\tilde{\bm{b}},\tilde{\bm{b}}^{\dagger}\right)^T$ are column vectors of the original and transformed bosonic operators respectively. The matrix $\bm{M}$ is a symplectic matrix that ensures that the new bosonic operators $\tilde{\bm{\xi}}$ satisfy the canonical commutation relations and $\bm{\gamma}$ is a vector of displacements. The symplecticity of $\bm{M}$ implies that $\bm{U}\bm{V}^T = \bm{V}\bm{U}^T$ and $\bm{U}\bm{U}^{\dagger} - \bm{V}\bm{V}^{\dagger} = \bm{1}$. The unitary operator $\mathcal{U}_{b}$ that performs the Bogoliubov transform, 
\begin{align}
    \tilde{\bm{\xi}} = \mathcal{U}_{b}^{\dagger}\,\bm{\xi}\, \mathcal{U}_{b} \equiv \bm{M}\bm{\xi} + \bm{\Gamma}
\end{align}
can be partitioned into two components $\mathcal{U}_{b} = \mathcal{U}_{d}\, \mathcal{U}_{q}$. The first component, performed by $\mathcal{U}_{q}$, mixes the original bosonic modes,
\begin{align}
    \bm{\xi} \to \bm{\bar{\xi}}  = \bm{M}\bm{\xi}.
\end{align} 
The second component, performed by $ \mathcal{U}_{d}$, displaces the mixed modes, 
\begin{align}
    \bm{\bar{\xi}} \to \tilde{\bm{\xi}} = \bm{\bar{\xi}} + \bm{\Gamma} = \bm{M}\bm{\xi} + \bm{\Gamma}.
\end{align}
  
The first component of the Bogoliubov transform that mixes the modes can be performed  as,
\begin{align}
    \bm{\bar{\xi}}  = \mathcal{U}_q^{\dagger}\,\bm{\xi}\,\mathcal{U}_q =  \bm{M}\bm{\xi}, \label{eq:bog_quad}
\end{align} 
using the quadratic unitary $\mathcal{U}_q$. To derive the form of $\mathcal{U}_q$, we use the equation of motion method \cite{wagner1986}. Let $\mathcal{U}_q(t) = e^{Xt}$ where 
\begin{align}
    X = \sum_{p,q=1}^{N} \alpha_{pq} {b}_p^{\dagger}{b}_q + \frac{1}{2}\left(\beta_{pq}{b}_p^{\dagger}{b}_q^{\dagger} + \beta_{pq}^{*}{b}_p{b}_q\right)
\end{align}
is the quadratic generator of the unitary transform and $X^{\dagger} = - X$ is ensured by $\bm{\alpha}^{\dagger} = -\bm{\alpha}$ and $\bm{\beta}^T = \bm{\beta}$. The transformed time-dependent operators are defined as $\bm{\bar{\xi}}(t)  = \mathcal{U}_q^{\dagger}(t)\,\bm{\xi}\,\mathcal{U}_q(t)$, with the initial conditions $\bm{\bar{\xi}}(0) = \bm{\xi}$ and we require $\bm{\bar{\xi}}(1) = \bm{M}\bm{\xi} $. Using canonical commutation relations, the equation of motion for $\bm{\bar{\xi}}(t)$ can be shown to be, 
\begin{align}
    \frac{d\bm{\bar{\xi}}(t)}{dt} = e^{-Xt}[\bm{\xi},X]e^{Xt} = \bm{\theta} \bm{\bar{\xi}}(t), \label{eq:xi_eom}
\end{align}
where the matrix $\bm{\theta}$ is defined as, 
\begin{align}
    \bm{\theta} \equiv \left(\begin{array}{cc}
    \bm{\alpha} & \bm{\beta} \\
    \bm{\beta}^{*}  & \bm{\alpha}^{*} \\
    \end{array}\right). 
\end{align}
Eq.~\eqref{eq:xi_eom} implies that $\bm{\bar{\xi}}(t) = e^{\bm{\theta}t}\bm{\bar{\xi}}(0) = e^{\bm{\theta}t} \bm{\xi}$. This satisfies our initial condition $\bm{\bar{\xi}}(0) = \bm{\xi}$. Furthermore, setting $t=1$, we obtain, $\bm{\bar{\xi}}(1) = e^{\bm{\theta}}\bm{\xi} =  \bm{M}\bm{\xi}$. Thus $\mathcal{U}_q = e^{X}$ satisfies Eq.~\eqref{eq:bog_quad} if $\bm{\theta} = \log{\bm{M}}$, and the form of $\bm{\theta}$ ensures that $\bm{M}$ is a symplectic matrix. This demonstrates that $\mathcal{U}_q = e^{X}$ mixes the bosonic modes as required in the first step on the Bogoliubov transform.

The second step of the Bogoliubov transform which displaces the bosonic modes $\tilde{\bm{\xi}} = \bm{\bar{\xi}} + \bm{\Gamma}$ is implemented by the displacement unitary operator 
\begin{align}
    \mathcal{U}_{d}(\bm{\xi}) &= e^{\sum_{p=1}^{N} \gamma_p b_p^{\dagger}- \gamma^{*}_p b_p} \label{eq:u_d}
\end{align}
Here we wish to emphasize that this displacement operator is a function of the original bosonic variables $\bm{\xi}$, as highlighted by the dependence indicated in the parenthesis: $\mathcal{U}_d(\bm{\xi})$. The full Bogoliubov unitary operator then becomes $\mathcal{U}_{b}(\bm{\xi}) = \mathcal{U}_{d}(\bm{\xi}) \mathcal{U}_{q}(\bm{\xi})$. Its action on the bosonic operators $\bm{\xi}$ can be demonstrated as, 
\begin{align}
    \tilde{\bm{\xi}} & = \mathcal{U}_{b}(\bm{\xi})^{\dagger}\, \bm{\xi}\,\mathcal{U}_{b}(\bm{\xi}) \\
    &= \mathcal{U}_{q}^{\dagger}(\bm{\xi}) \mathcal{U}_{d}^{\dagger}(\bm{\xi})\,\bm{\xi}\,\mathcal{U}_{d}(\bm{\xi})\mathcal{U}_{q}(\bm{\xi}) \\
    & = \Bigl[\mathcal{U}_{q}^{\dagger}(\bm{\xi}) \mathcal{U}_{d}^{\dagger}(\bm{\xi})\mathcal{U}_{q}(\bm{\xi})\Bigr]\Bigl[\mathcal{U}_{q}^{\dagger}(\bm{\xi})\,\bm{\xi}\,\mathcal{U}_{q}(\bm{\xi})\Bigr] \notag \\
    & \times \Bigl[\mathcal{U}_{q}^{\dagger}(\bm{\xi})\mathcal{U}_{d}(\bm{\xi})\mathcal{U}_{q}(\bm{\xi})\Big] \label{eq:bog_mix} \\
    & = \mathcal{U}_{d}^{\dagger}(\bm{\bar{\xi}})\,\bar{\bm{\xi}}\,\mathcal{U}_{d}(\bm{\bar{\xi}}) \label{eq:bog_disp} \\
    & = \bm{\bar{\xi}} + \bm{\Gamma} = \bm{M}\bm{\xi} + \bm{\Gamma} \label{eq:full_bog}.
\end{align}
Here we used the form of composition of two unitary transforms recommended by Schwinger \cite{schwinger1960,liu2024}. In Eq.~\eqref{eq:bog_mix}, the original bosonic modes $\bm{\xi}$ are mixed into new modes $\bm{\bar{\xi}}$ in the second parenthesis, whereas the displacement unitary is rotated into the basis of these new modes in the first and third parenthesis. This is highlighted in Eq.~\eqref{eq:bog_disp} where the displacement unitaries are explicitly a function of the mixed modes $\bm{\bar{\xi}}$. Equation~\eqref{eq:full_bog} highlights the effect of the displacement, resulting in the mixed and displaced modes $\tilde{\bm{\xi}} = \bm{M}\bm{\xi} + \bm{\Gamma}$.

\section{Bloch-Messiah Decomposition of the Bogoliubov Unitary} \label{app:unr_bog}

The symplectic Bogoliubov matrix $\bm{M}$ can be decomposed into a product of two unitary matrices, $\bm{M}_1$ and $\bm{M}_3$, and a non-negative diagonal symplectic matrix, $\bm{M}_2$, using the Bloch - Messiah decomposition \cite{braunstein2005,houde2024},
\begin{align}
    \bm{M} &= \left(\begin{array}{cc}
    \bm{U} & -\bm{V} \\
    -\bm{V}^{*} &  \bm{U}^{*} \\
    \end{array}\right) \notag \\ 
    & = \left(\begin{array}{cc}
    \bm{W} & \bm{0} \\
    \bm{0} &  \bm{W}^{*} \\
    \end{array}\right) \left(\begin{array}{rr}
    \bm{U}_D & -\bm{V}_D \\
    -\bm{V}_D &  \bm{U}_D \\
    \end{array}\right) \left(\begin{array}{cc}
    \bm{X}^{\dagger} & \bm{0} \\
    \bm{0} &  \bm{X}^{T} \\
    \end{array}\right)  \\
    & = \bm{M}_3 \bm{M}_2 \bm{M}_1. \label{eq:m_mats}
\end{align}
These matrices can be obtained using the singular value decomposition, $\bm{U} = \bm{W}\bm{U}_D\bm{X}^{\dagger}$ and $\bm{V} = \bm{W}\bm{V}_D\bm{X}^{T}$, with $\bm{W}$ and $\bm{X}$ being unitary matrices, and $\bm{U}_D$ and $\bm{V}_D$ being diagonal matrices with positive singular values, satisfying $\bm{U}_D^2 - \bm{V}_D^2 = \bm{1}$.
Equation~\eqref{eq:m_mats} defines the matrices $\bm{M}_1, \bm{M}_2$, and $\bm{M}_3$. The Bogoliubov transform represented by the matrix $\bm{M}$ can thus be decomposed into the three transforms represented by these matrices. For the Bogoliubov transforms considered in this work, the matrix $\bm{M}$ is real, resulting in real orthogonal matrices $\bm{W}, \bm{X}, \bm{M}_1$ and $\bm{M}_3$. The transformation corresponding to $\bm{M}_1$ mixes the creation operators among themselves, and similar for the annihilation operators, and is similar to orbital rotations in electronic structure theory. It can be represented using a unitary transform as $\bm{M}_1 \bm{\xi} = \mathcal{U}_1^{\dagger}\,\bm{\xi}\,\mathcal{U}_1 $, with 
\begin{align}
    \mathcal{U}_1 & = e^{\sum_{p>q=1}^{N}  \left(\log{\bm{X}^{\dagger}}\right)_{pq} \left({b}_p^{\dagger}{b}_q - {b}_q^{\dagger}{b}_p\right)} \notag \\
    & = \prod_{p>q=1}^{N} e^{\chi_{pq} \left({b}_p^{\dagger}{b}_q - {b}_q^{\dagger}{b}_p\right)} = \prod_{p>q=1}^{N} BS_{p,q}\left(2\chi_{pq},\frac{\pi}{2}\right). \label{eq:bs_unit}
\end{align}
The parameters $\chi_{pq}$ can be obtained from $\left(\log{\bm{X}^{\dagger}}\right)_{pq}$   \cite{reck1994}. The last equality in Eq.~\eqref{eq:bs_unit} denotes that the unitary $\mathcal{U}_1$ can be partitioned into a product of elementary beam-splitter gates defined in Eq.~\eqref{eq:u_bs}. Similar arguments hold for the transform corresponding to the matrix $\bm{M}_3$, resulting in, 
\begin{align}
    \mathcal{U}_3 = \prod_{p>q=1}^{N} BS_{p,q}\left(2\phi_{pq},\frac{\pi}{2}\right),
\end{align}
where the rotation angles $\phi_{pq} $ can be obtained from $\left(\log{\bm{W}}\right)_{pq}$. Furthermore, the transform corresponding to the matrix $\bm{M}_2$ is performed by the unitary operator,
\begin{align}
    \mathcal{U}_2 &= e^{\sum_{p=1}^{N} \zeta_p \left({b_p^{\dagger}}^2 - b_p^2\right)} \notag \\
    & = \prod_{p=1}^{N}  e^{ \zeta_p \left({b_p^{\dagger}}^2 - b_p^2\right)} = \prod_{p=1}^{N} S_p\left(\zeta_p\right), \label{eq:sq_unit}
\end{align}
with $\cosh(\bm{\zeta}) = \bm{U}_D$,  $\sinh(\bm{\zeta}) = -\bm{V}_D$, $\bm{\zeta}$ is a diagonal matrix of $\{\zeta_p\}$, and the last equality in Eq.~\eqref{eq:sq_unit} highlights that the unitary $\mathcal{U}_2$ can be decomposed into elementary one-mode squeezing gates defined in Eq.~\eqref{eq:u_sq}. Lastly, the displacement unitary can be decomposed into elementary displacement gates defined in Eq.~\eqref{eq:u_disp}, as
\begin{align}
    \mathcal{U}_d = \prod_{p=1}^{N} \mathcal{D}_p(\gamma_p).
\end{align}
Combining all these elements, the Bogoliubov unitary can be written as $\mathcal{U}_b = \mathcal{U}_d \, \mathcal{U}_q = \mathcal{U}_d \, \mathcal{U}_3 \, \mathcal{U}_2 \, \mathcal{U}_1 $. This is highlighted in Eq.~\eqref{eq:bog_u_decomp}.

\end{document}